\documentclass[twocolumn]{aastex701}

\begin{document}

\title{The Influence of Hubble Expansion on the Future Merger of the Milky Way and Andromeda}

\author[0009-0006-0194-6211]{Yufan Fane Zhou}
%\email[show]{yufanz@smail.nju.edu.cn}
\email{yufanz@smail.nju.edu.cn}
\affiliation{School of Astronomy and Space Science, Nanjing University, Nanjing 210046, China}
\affiliation{Key Laboratory of Modern Astronomy and Astrophysics (Nanjing University), Ministry of Education, Nanjing 210046, China}

\author[0000-0001-8101-8293]{Tianchi Zhang}
\email{tczhang@bjp.org.cn}
\affiliation{Beijing Planetarium, Beijing Academy of Science and Technology, Beijing 100044, China}

\author[0000-0001-7075-6098]{Shihong liao}
\email{shliao@nao.cas.cn}
\affiliation{National Astronomical Observatories, Chinese Academy of Sciences, Beijing 100101, China}

\author[0000-0003-0355-6437]{Zhiyuan Li}
\email{lizy@nju.edu.cn}
\affiliation{School of Astronomy and Space Science, Nanjing University, Nanjing 210046, China}
\affiliation{Key Laboratory of Modern Astronomy and Astrophysics (Nanjing University), Ministry of Education, Nanjing 210046, China}
\affiliation{Institute of Science and Technology for Deep Space Exploration, Suzhou Campus, Nanjing University, Suzhou 215163, China}

\author[0000-0002-9788-2577]{Lan Wang}
\email{wanglan@bao.ac.cn}
\affiliation{National Astronomical Observatories, Chinese Academy of Sciences, Beijing 100101, China}

\author[0000-0002-9578-3081]{David Benisty}
\email{david.benisty@biu.ac.il}
\affiliation{Department of Physics, Bar-Ilan University, 5290002 Ramat Gan, Israel}

\begin{abstract}
Most existing idealized simulations of galaxy mergers neglect Hubble expansion, as the local Hubble flow is curved by overdensities and the expansion effect is weak at these scales. However, even in the gravity-dominated Local Group (LG), the expansion itself still persists, and it has been shown important in some LG-related studies. Here we take the future merger of the Milky Way (MW) and Andromeda (M31) as a case study to investigate the differences between simulations with and without Hubble expansion. We perform a suite of idealized \textit{N}-body simulations of the MW--M31 system under pure Newtonian, $\Lambda$CDM, and $w_0w_a$CDM scenarios, testing different masses and proper motions of galaxies informed by current observational constraints. 
At low M31 tangential velocities, especially in cases where the two galaxies merge within a Hubble time, the expansion effect is negligible; while at high velocities ($>66.5\,\mathrm{km\,s^{-1}}$), the $\Lambda$CDM simulations introduce a noticeable delay in the merger time compared to the Newtonian ones and may even alter the final fate of the system, but the caveat here is that the isolated-pair idealization and the assumed expansion history break down.
The delay is also present but less pronounced in the $w_0w_a$CDM runs. When M33 and the Large Magellanic Cloud are included, the differences in merger times are reduced. 
In summary, Hubble expansion can play a noticeable role for galaxy mergers with high tangential velocities, but its actual influence on the MW--M31 merger depends on the precise masses and proper motions to be determined in the future.
\end{abstract}



\section{Introduction}
\label{sec:introduction}
Almost all numerical simulations of galaxy mergers, including studies on the future merger of the Milky Way (MW) and the Andromeda Galaxy (M31), have not taken Hubble expansion into account \citep{dubinski1996,cox2008,vandermarel2012,vandermarel2019,schiavi2020,sawala2025,wu2026}. This is based on the fact that the overdensity of the Local Group (LG) curves the local Hubble flow \citep{sandage1986,karachentsev2009,karachentsev2013}, or on the assumption that the expansion effect is negligible on the scale of galaxy pairs.

However, the deviation from the Hubble flow within the LG merely means the dominance of gravity on that scale; it does not imply that cosmic expansion is absent. Since \citet{mcvittie1933} introduced a spacetime metric describing a point mass embedded in a Friedmann-Lemaitre-Robertson-Walker universe, the influence of Hubble expansion on gravitationally bound systems has been widely studied \citep[e.g.,][]{sereno2007tt,faraoni2007,nandra2012ug,benisty2025mfv}.

Recent studies \citet{karachentsev2009,penarrubia2014,benisty2021,benisty2022,benisty2024} find that considering expansion in the LG is important, because embedding the cosmological constant $\Lambda$ alone or the full $\Lambda$CDM expansion history significantly changes the LG mass predicted by the \textit{timing argument} \citep{kahn1959,lynden-bell1981,li2008,strigari2025}. In this paper, we aim to investigate whether Hubble expansion can affect the future of the LG. We perform idealized \textit{N}-body simulations including the MW and M31 with or without Hubble expansion, and compare the differences in merger time between them.

Current constraints on the masses of the MW and M31 remain uncertain: MW mass measurements range from $\sim0.5\times10^{12}\,M_{\odot}$ to $\sim2\times10^{12}\,M_{\odot}$ \citep{wang2020}, while M31 mass measurements range from a few times $10^{11}\,M_{\odot}$ to $\sim3\times10^{12}\,M_{\odot}$ \citep{bhattacharya2025}. Furthermore, different studies yield different proper motions for M31 \citep[e.g.,][]{sohn2012,vandermarel2019,salomon2021,wu2025}, corresponding to tangential velocities ranging from $\sim0\,\mathrm{km\,s^{-1}}$ to $>100\,\mathrm{km\,s^{-1}}$. Here we do not aim to provide the full possible range of solutions for the future of the MW--M31 system, as that would require substantial effort to scan the plausible ranges of parameters such as galaxy masses, positions, and motions, much like what \citet{sawala2025,wu2026} did. The main purpose of this work is to illustrate the influence of Hubble expansion on the future merger of the MW and M31, and to determine whether it is negligible or not.

The rest of the paper is structured as follows. Section~\ref{sec:methods} describes the method of the idealized simulations, including the \textit{N}-body code we used, the implementation of expansion, the galaxy models, and the simulation setup. The results are shown in Section~\ref{sec:results}. Section~\ref{sec:m33lmc} provides several supplementary simulations that include M33 and the Large Magellanic Cloud (LMC). A brief conclusion is presented in Section~\ref{sec:conclusion}.

\section{Methods}
\label{sec:methods}
We construct the \textit{N}-body initial conditions for the MW and M31, and then perform idealized simulations both with and without Hubble expansion using the numerical simulation code \textsc{gadget-4} \citep{springel2021}. Two different MW masses, two different M31 masses, three different values of M31's proper motion in the right ascension direction, and three different values of M31's proper motion in the declination direction are adopted, all within the range of current observational uncertainties. In addition, for the simulations including expansion, we test not only the standard $\Lambda$CDM model but also the $w_0w_a$CDM model favored by recent DESI DR2 results \citep{abdulKarim2025}.

\subsection{Simulation Code}
\label{subsec:gadget4}
\textsc{gadget-4} \citep{springel2021} is a massively parallel tree-based \textit{N}-body code that supports a variety of integration modes and boundary conditions. The code can run Newtonian simulations in physical coordinates with vacuum boundary conditions, which is the classic setup for idealized galaxy encounters.

Alternatively, it can perform cosmological integrations in comoving coordinates while still employing vacuum boundaries rather than periodic boxes. In this mode, \textsc{gadget-4} solves the Friedmann equation for a $\Lambda$CDM cosmology with parameters $\Omega_{\rm m}$, $\Omega_\Lambda$, and $H_0$ specified in the parameter file and evolves particles in comoving coordinates, enabling the study of galaxy evolution with Hubble expansion.

Compared to $\Lambda$CDM, results from DESI DR2 \citep{abdulKarim2025} favor a dynamical dark energy model (also referred to as $w_0w_a$CDM), which adopts the parameterization $w(a)=w_{0}+w_{a}(1-a)$. Therefore, in our simulations with cosmic expansion, we study not only the $\Lambda$CDM scenario but also the $w_0w_a$CDM scenario.

In the source code of \textsc{gadget-4}, the dimensionless Hubble parameter $E(a)=H(a)/H_{0}$ for a $\Lambda$CDM cosmology is implemented as
\begin{equation}
E(a) = \sqrt{\Omega_{\rm m}\,a^{-3} + \Omega_{k}\,a^{-2} + \Omega_{\Lambda}} .
\end{equation}
To describe a $w_0w_a$CDM cosmology, we modify this expression to \citep[e.g.,][]{abdulKarim2025}
\begin{equation}
E(a) = \sqrt{\Omega_{\rm m}a^{-3} + \Omega_{k}a^{-2} + \Omega_{\Lambda} a^{-3(1+w_{0}+w_{a})} e^{-3w_{a}(1-a)}} .
\end{equation}
In our simulations, the curvature density parameter is $\Omega_{k}=0$ (assuming a flat universe). Values of the matter density parameter $\Omega_{\rm m}$, the dark energy density parameter $\Omega_{\Lambda}$, the present-day dark energy equation-of-state parameter $w_{0}$, and its time variation $w_{a}$ are summarized in Section~\ref{subsec:settings}.

\subsection{Initial Galaxy Models}
\label{subsec:model}

\begin{table*}
\centering
\caption{36 different galaxy mass and proper motion value combinations across all 108 MW--M31 merger simulations.}
\label{tab:settings}
\begin{tabular}{lcccccc}
\hline
Label & $M_{\rm MW}\ (10^{10}\,M_{\odot})$ & $M_{\rm M31}\ (10^{10}\,M_{\odot})$ & $\mu_{\rm \alpha*,M31}\ (\mu\mathrm{as\,yr^{-1}})$ & $\mu_{\rm \delta,M31}\ (\mu\mathrm{as\,yr^{-1}})$ & $v_{\rm t,M31}\ (\mathrm{km\,s^{-1}})$ \\
\hline
\verb|m80m100_378-271| & 80 & 100 & 37.8 & -27.1 & 26.0 \\
\verb|m80m100_378-205| & 80 & 100 & 37.8 & -20.5 & 6.3 \\
\verb|m80m100_378-139| & 80 & 100 & 37.8 & -13.9 & 24.6 \\
\verb|m80m100_459-271| & 80 & 100 & 45.9 & -27.1 & 44.3 \\
\verb|m80m100_459-205| & 80 & 100 & 45.9 & -20.5 & 36.4 \\
\verb|m80m100_459-139| & 80 & 100 & 45.9 & -13.9 & 43.5 \\
\verb|m80m100_540-271| & 80 & 100 & 54.0 & -27.1 & 71.1 \\
\verb|m80m100_540-205| & 80 & 100 & 54.0 & -20.5 & 66.5 \\
\verb|m80m100_540-139| & 80 & 100 & 54.0 & -13.9 & 70.6 \\
\verb|m80m160_378-271| & 80 & 160 & 37.8 & -27.1 & 26.0 \\
\verb|m80m160_378-205| & 80 & 160 & 37.8 & -20.5 & 6.3 \\
\verb|m80m160_378-139| & 80 & 160 & 37.8 & -13.9 & 24.6 \\
\verb|m80m160_459-271| & 80 & 160 & 45.9 & -27.1 & 44.3 \\
\verb|m80m160_459-205| & 80 & 160 & 45.9 & -20.5 & 36.4 \\
\verb|m80m160_459-139| & 80 & 160 & 45.9 & -13.9 & 43.5 \\
\verb|m80m160_540-271| & 80 & 160 & 54.0 & -27.1 & 71.1 \\
\verb|m80m160_540-205| & 80 & 160 & 54.0 & -20.5 & 66.5 \\
\verb|m80m160_540-139| & 80 & 160 & 54.0 & -13.9 & 70.6 \\
\verb|m100m100_378-271| & 100 & 100 & 37.8 & -27.1 & 26.0 \\
\verb|m100m100_378-205| & 100 & 100 & 37.8 & -20.5 & 6.3 \\
\verb|m100m100_378-139| & 100 & 100 & 37.8 & -13.9 & 24.6 \\
\verb|m100m100_459-271| & 100 & 100 & 45.9 & -27.1 & 44.3 \\
\verb|m100m100_459-205| & 100 & 100 & 45.9 & -20.5 & 36.4 \\
\verb|m100m100_459-139| & 100 & 100 & 45.9 & -13.9 & 43.5 \\
\verb|m100m100_540-271| & 100 & 100 & 54.0 & -27.1 & 71.1 \\
\verb|m100m100_540-205| & 100 & 100 & 54.0 & -20.5 & 66.5 \\
\verb|m100m100_540-139| & 100 & 100 & 54.0 & -13.9 & 70.6 \\
\verb|m100m160_378-271| & 100 & 160 & 37.8 & -27.1 & 26.0 \\
\verb|m100m160_378-205| & 100 & 160 & 37.8 & -20.5 & 6.3 \\
\verb|m100m160_378-139| & 100 & 160 & 37.8 & -13.9 & 24.6 \\
\verb|m100m160_459-271| & 100 & 160 & 45.9 & -27.1 & 44.3 \\
\verb|m100m160_459-205| & 100 & 160 & 45.9 & -20.5 & 36.4 \\
\verb|m100m160_459-139| & 100 & 160 & 45.9 & -13.9 & 43.5 \\
\verb|m100m160_540-271| & 100 & 160 & 54.0 & -27.1 & 71.1 \\
\verb|m100m160_540-205| & 100 & 160 & 54.0 & -20.5 & 66.5 \\
\verb|m100m160_540-139| & 100 & 160 & 54.0 & -13.9 & 70.6 \\
\hline
\end{tabular}
\vspace{2mm}

\raggedright
\footnotesize \textbf{Notes.} For each parameter combination, three simulations (Newtonian, $\Lambda$CDM, and $w_0w_a$CDM) are run separately, resulting in a total of 108 simulations.
\end{table*}

We construct the initial conditions of the MW and M31 using the method described in \citet{springel2005ic}. The dark matter halo of each galaxy is modeled as a Hernquist \citep{hernquist1990} profile matched to a Navarro--Frenk--White \citep[NFW, ][]{navarro1996,navarro1997} profile. We also place a massive black hole (BH) particle at the center of each galaxy, which serves as a convenient marker for the galactic center when computing the separation between the two galaxies.

The fiducial virial masses are $M_{\rm vir}=1.0\times10^{12}\,M_{\odot}$ for the MW and $M_{\rm vir}=1.6\times10^{12}\,M_{\odot}$ for M31, and the halo concentration of them are both $c=12$ \citep{klypin2002}. Dark matter particles have a mass of $5\times10^{5}\,M_{\odot}$, while the central BH particle has a mass of $5\times10^{7}\,M_{\odot}$ in both galaxies. The total particle numbers are approximately $2.0\times10^{6}$ for the MW and $3.2\times10^{6}$ for M31. In addition, we constructed a low-mass MW model with $M_{\rm vir}=8.0\times10^{11}\,M_{\odot}$ and a low-mass M31 model with $M_{\rm vir}=1.0\times10^{12}\,M_{\odot}$, for which the particle numbers are correspondingly reduced.

\citet{cox2008} and \citet{schiavi2020} also included an intergalactic medium (IGM) component in their simulations, because early estimates of the LG mass significantly exceeded the sum of the MW and M31. However, as \citet{vandermarel2012} pointed out, more recent estimates \citep[see also][]{benisty2024} yield a LG mass nearly equal to the sum of the two galaxies, making the inclusion of IGM less necessary. Furthermore, following similar works \citep[e.g.,][]{sawala2025,wu2026}, we model our galaxies without disk and bulge components, since the merger's orbital evolution is dominated by the significantly more massive dark matter halos. Through numerical tests including these components, we confirmed that their effect on the merger process reported here is negligible (see Appendix~\ref{sec:appendix}).

\subsection{Coordinates and Velocities of M31}
\label{subsec:galactocentric}
As mentioned in Section~\ref{sec:introduction}, the possible range of M31's proper motion values is wide, and \citet{salomon2021} finds that the measurements based on Gaia's blue and red star samples exhibit large discrepancies. \citet{wu2025} points out that this inconsistency originates from systematic differences between Gaia's five-parameter (5p) and six-parameter (6p) astrometric solutions, and that the reliability of the 6p solution is limited. After excluding extreme red sources and calibrating the proper motion zero-points separately for 5p and 6p sources, \citet{wu2025} provides a robust Gaia-based result: $(\mu_{\alpha*},\mu_{\delta})_\mathrm{M31}=(45.9\pm8.1,-20.5\pm6.6)\,\mathrm{\mu as\,yr^{-1}}$.

Therefore, based on \citet{wu2025}, we adopt $\mu_{\alpha*}=37.8$, $45.9$, and $54.0\,\mu\mathrm{as\,yr^{-1}}$ and $\mu_{\delta}=-27.1$, $-20.5$, and $-13.9\,\mu\mathrm{as\,yr^{-1}}$, yielding nine proper-motion combinations in total. For the other parameters of M31, we adopt RA = $00^{\rm h}42^{\rm m}44.3^{\rm s}$, Dec = $+41^{\rm d}16^{\rm m}09^{\rm s}$ \citep{mcconnachie2012}, a distance of $D = 784\,\mathrm{kpc}$ \citep{stanek1998}, and a line-of-sight velocity of $V_{\rm LOS} = -301.0\,\mathrm{km\,s^{-1}}$ \citep{watkins2013}, consistent with \citet{wu2025}.

To convert the coordinates and velocities of M31 into the Galactocentric frame for the convenience of the simulation setup, we adopt the Galactocentric distance of the Sun of $R_{0}=8.122\,\mathrm{kpc}$ \citep{gravity2018}, the solar motion of $(U_\odot, V_\odot, W_\odot) = (7.01, 10.13, 4.95)\,\mathrm{km\,s^{-1}}$ \citep{huang2015}, and the circular speed at the solar position of $V_{\rm c}(R_{0}) = 234.04\,\mathrm{km\,s^{-1}}$ \citep{zhou2023}.

Based on the above parameters, the nine proper-motion combinations yield M31 tangential velocities ranging from $6.3$ to $71.1\,\mathrm{km\,s^{-1}}$, as listed in the last column of Table~\ref{tab:settings}.

\subsection{Simulation Settings}
\label{subsec:settings}

The two MW masses and the two M31 masses in Section~\ref{subsec:model}, together with the nine proper motion values in Section~\ref{subsec:galactocentric}, yield 36 parameter combinations in total (see Table~\ref{tab:settings}). For each combination, we run one Newtonian, one $\Lambda$CDM, and one $w_0w_a$CDM simulation, respectively. The simulations are labelled as, e.g., \verb|N_m100m160_540-139|, where the leading letter N, L, or W denotes the simulation type (Newtonian, $\Lambda$CDM, or $w_0w_a$CDM), and the rest of the name specifies the galaxy masses and the M31 proper-motion parameters (consistent with Table~\ref{tab:settings}).

(a) Newtonian simulation: Galaxy particles evolve under gravitational force. The simulation is carried out in physical coordinates, starting at $t=0$ and integrating forward in time.

\begin{figure*}
    \centering
    \includegraphics[width=0.9\linewidth]{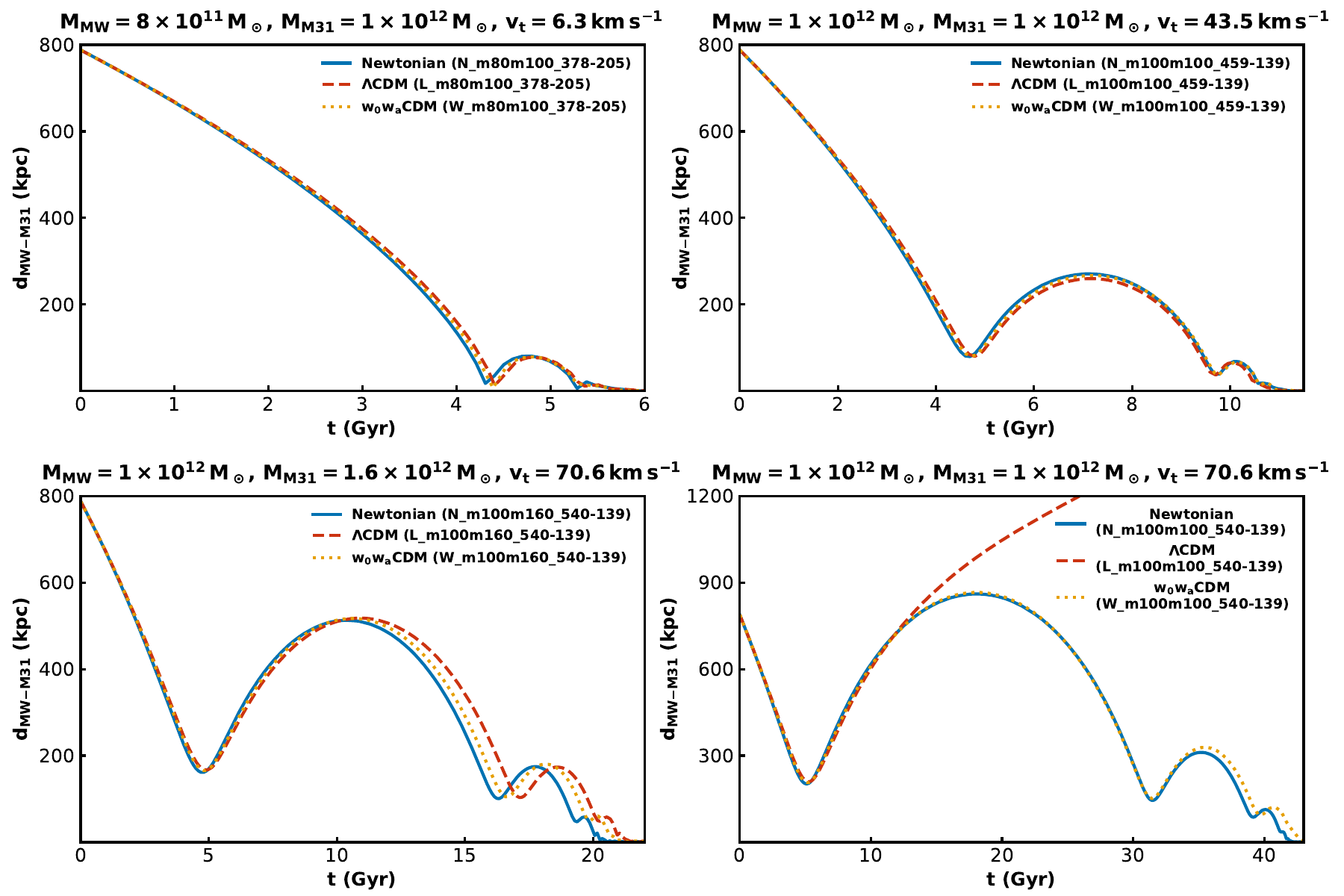}
    \caption{Distance--time curves of 12 selected representative MW--M31 merger simulations. The parameters of the simulations are given in the title and legend of each panel. Line styles distinguish the cosmology: Newtonian (solid), $\Lambda$CDM (dashed), and $w_0w_a$CDM (dotted).}
    \label{fig:idealized}
\end{figure*}

\begin{figure*}
    \centering
    \includegraphics[width=0.8\linewidth]{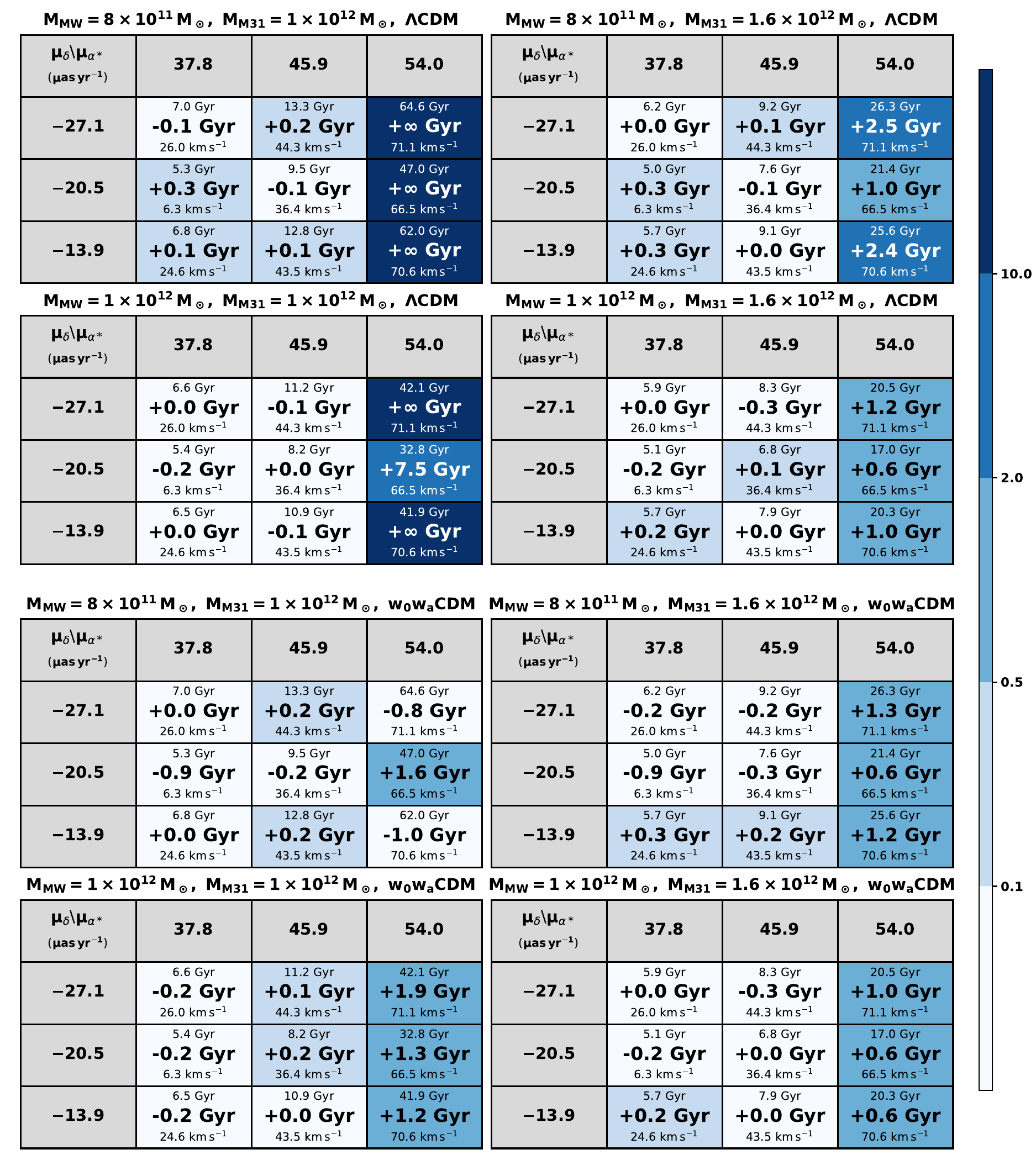}
    \caption{Upper four panels: The merger time differences between the 36 $\Lambda$CDM simulations and their 36 Newtonian counterparts. The title of each sub-panel gives the galaxy masses, and the first row and the first column give the adopted M31 proper motions, while the remaining nine cells show the results. In each result cell, the three rows show, from top to bottom, the merger time of the Newtonian simulation $t_{\rm m,N}$; the merger time difference between the $\Lambda$CDM and Newtonian simulations, $\Delta t = t_{\rm m,L} - t_{\rm m,N}$; and the M31 tangential velocity of the simulation, respectively. The value $+\infty$ indicates that the MW and M31 do not merge in the $\Lambda$CDM simulation of that cell. The background colors from light to dark correspond to $\Delta t < 0.1\,\mathrm{Gyr}$, $0.1\,\mathrm{Gyr} \le \Delta t < 0.5\,\mathrm{Gyr}$, $0.5\,\mathrm{Gyr} \le \Delta t < 2.0\,\mathrm{Gyr}$, $2.0\,\mathrm{Gyr} \le \Delta t < 10.0\,\mathrm{Gyr}$, and $\Delta t \ge 10.0\,\mathrm{Gyr}$. Lower four panels: The merger time differences between the 36 $w_0w_a$CDM simulations and their 36 Newtonian counterparts.}
    \label{fig:mergertime}
\end{figure*}

(b) $\Lambda$CDM simulation: The cosmological parameters are $\Omega_{\rm m}=0.315$, $\Omega_{\Lambda}=0.685$, and $H_{0}=67.4\,\mathrm{km\,s^{-1}\,Mpc^{-1}}$ \citep{planck2020}. In this case, the calculated velocity of M31 in the Galactocentric frame is understood as the total velocity, which consists of a peculiar velocity and a Hubble velocity:
\begin{equation}
v_{\rm rad} = v_{\rm pec} + v_{\rm H},
\end{equation}
where $v_{\rm H} = H_{0}D$ is naturally accounted for by the expanding background. For example, in \verb|N_m100m160_540-139|, the position and velocity of M31 are $(-385.5, 623.8, -288.3)\,\mathrm{kpc}$ and $(-6.9, -116.7, 60.8)\,\mathrm{km\,s^{-1}}$, respectively. The Hubble velocity is $v_{\rm H}=(-26.0, 42.0, -19.4)\,\mathrm{km\,s^{-1}}$, so the peculiar velocity we set in the initial condition file of \verb|L_m100m160_540-139| is $(19.1, -158.7, 80.2)\,\mathrm{km\,s^{-1}}$. The simulation is carried out in comoving coordinates with vacuum boundary conditions, starting at scale factor $a=1$ and integrating forward.

(c) $w_0w_a$CDM simulation: The cosmological parameters are $\Omega_{\rm m}=0.319$, $\Omega_{\Lambda}=0.681$, $H_{0}=66.7\,\mathrm{km\,s^{-1}\,Mpc^{-1}}$, $w_{0}=-0.752$, and $w_{a}=-0.86$ \citep{abdulKarim2025}. Here, the peculiar velocities in the initial condition files are corrected in the similar way as in the $\Lambda$CDM scenario. The simulation is also carried out in comoving coordinates with vacuum boundary conditions.

In all simulations, the integration accuracy and the opening angle are $\mathrm{ErrTolIntAccuracy=0.012}$ and $\mathrm{ErrTolTheta=0.6}$, respectively. A constant physical softening length of $0.5\,\mathrm{kpc}$ is adopted across all Newtonian, $\Lambda$CDM, and $w_0w_a$CDM simulations. We consider the MW and M31 to be merged when their separation decreases to 10$\,\mathrm{kpc}$. All simulations are run at least until the moment of merger (for some $\Lambda$CDM simulations in which the galaxies do not merge, the simulation durations are the same as their corresponding Newtonian ones).

In \textsc{gadget-4} integrations in comoving coordinates with vacuum boundary conditions, the equation of motion carries a background-density correction term (eq~(55) of \citet{springel2001}). When the background correction is combined with the $\ddot a/a$ term of the Friedmann equation, the matter contributions cancel exactly, and transforming to physical coordinates leaves Newtonian gravity plus a repulsive central acceleration proportional to the distance from the origin, $\Omega_\Lambda H_0^{2}\mathbf{r}$ for $\Lambda$CDM, or the corresponding time-dependent dark energy term for $w_0w_a$CDM. The Newtonian, $\Lambda$CDM, and $w_0w_a$CDM simulations, which adopt the same initial conditions, therefore differ from one another by this single central force term only. Its amplitude grows linearly with the pair separation, so qualitatively the term is negligible while the two galaxies are close and becomes important for wide orbits, such as those with high M31 tangential velocities. The quantitative results are presented in Section~\ref{sec:results}.

\section{Results}
\label{sec:results}
Figure~\ref{fig:idealized} shows the distance--time curves of 12 selected representative simulations, covering the low, intermediate, and high M31 tangential velocities and illustrating the four main types of outcomes among all 108 simulations. The four panels correspond to four different mass and proper motion combinations, and the three Newtonian, $\Lambda$CDM, and $w_0w_a$CDM simulations in each panel are distinguished using solid, dashed, and dotted line styles. Figure~\ref{fig:mergertime} shows the merger time differences between the 36 $\Lambda$CDM simulations (or $w_0w_a$CDM simulations) and their 36 Newtonian counterparts, where darker colors indicate that the merger in the $\Lambda$CDM (or $w_0w_a$CDM) case lags behind the Newtonian case by a larger amount. The time differences in Figure~\ref{fig:mergertime} are not necessarily caused by the expansion; they may also arise from noise. To provide a noise floor, we first read off the merger-time differences resulting from different resolutions (see Figure~\ref{figs:lowresolution}), which is less than 0.2\,Gyr. Next, by changing the merger criterion from 10\,kpc to 20\,kpc, we find that the time differences in all simulations change by less than 0.3\,Gyr (except for the special case mentioned at the end of point (1) in Section~\ref{subsec:mergertime}). Therefore, we conservatively regard 0.3\,Gyr as the noise floor of the merger-time differences in Figure~\ref{fig:mergertime}.

\subsection{The First Pericenter}
\label{subsec:first}
In all simulations, the first pericenter occurs at approximately 4--5~Gyr (see Figure~\ref{fig:idealized}). Owing to the expansion effect, the $\Lambda$CDM and the $w_0w_a$CDM simulations reach the first pericenter slightly later, which is more clearly shown in the upper-left panel of Figure~\ref{fig:idealized} because of its smaller horizontal-axis range. In the upper two panels of Figure~\ref{fig:idealized}, since the M31 tangential velocity is not very high, the first pericenter distance is small, and thus the energy dissipation caused by dynamical friction before the first pericenter is relatively significant. Because the $\Lambda$CDM and $w_0w_a$CDM simulations reach the first pericenter later, their energy dissipation accumulates more, leading to a slightly smaller first pericenter distance than in the Newtonian simulations. In contrast, in the lower two panels, the first pericenter distance is larger and the energy dissipation effect is not so significant, so owing to the expansion effect the first pericenter distances in the $\Lambda$CDM and $w_0w_a$CDM simulations are slightly larger than in the Newtonian ones.

\subsection{The Final Merger Time}
\label{subsec:mergertime}
As for the differences in the final merger time between the simulations with and without expansion, the results can be divided into the following four categories.

(1) The expansion effect is negligible, and the merger time in the $\Lambda$CDM (or $w_0w_a$CDM) scenario shows only a barely noticeable delay relative to the Newtonian one. The upper-left panel of Figure~\ref{fig:idealized} serves as an example. Because of its low M31 tangential velocity, the separation between the MW and M31 decreases rapidly to small values, causing the expansion effect to be strongly suppressed, and the differences between the Newtonian and $\Lambda$CDM (or $w_0w_a$CDM) simulations at the first pericenter are subsequently erased by the later dynamical evolution. However, we note that Figure~\ref{fig:mergertime} shows that in this example the merger time of the $w_0w_a$CDM simulation is 0.9\,Gyr earlier than that of the Newtonian one. This is simply because, for this set of orbital parameters, the first pericenter distance is very close to the merger distance criterion (10\,kpc), and the galaxies are identified as merged once their separation happens to fall below this criterion.

(2) The expansion effect is negligible, and the merger time in the $\Lambda$CDM (or $w_0w_a$CDM) scenario is slightly earlier than in the Newtonian one. The upper-right panel of Figure~\ref{fig:idealized} serves as an example. This seemingly counterintuitive conclusion can be explained by our previous discussion of the first pericenter (see Section~\ref{subsec:first}). Since the expansion delays the time of the first pericenter, the galaxies remain near the pericenter for longer time than in the Newtonian simulations, and their energy is more strongly dissipated by dynamical friction. As a result, the distance at the next apocenter is smaller than in the Newtonian scenario, which in turn leads to an earlier final merger time.

(3) The expansion effect is not negligible, and the merger time in the $\Lambda$CDM (or $w_0w_a$CDM) scenario shows a noticeable delay relative to the Newtonian one. The lower-left panel of Figure~\ref{fig:idealized} serves as an example. In these simulations with high M31 tangential velocities, the separation rebounds to a significant value after the first pericenter, reaching $\sim 500$ kpc. The larger separation leads to a larger Hubble velocity, and the longer orbital timescale allows the expansion effect to accumulate more sufficiently. Figure~\ref{fig:mergertime} shows that in this example the merger time in the $\Lambda$CDM ($w_0w_a$CDM) scenario is 1\,Gyr (0.6\,Gyr) later than in the Newtonian one.

(4) The expansion effect is not negligible, and the MW and M31 eventually separate in the $\Lambda$CDM scenario. The lower-right panel of Figure~\ref{fig:idealized} serves as an example. This example demonstrates a qualitatively new result, showing that for low masses and high tangential velocities, the fate of the galaxy pair may be completely different depending on whether cosmic expansion is included or not. We see from the panel that in the Newtonian simulation the post-pericenter apocenter is about 800 kpc, a substantial fraction of the de Sitter radius $r_\Lambda=(\frac{GM}{\Omega_{\Lambda}H_{0}^{2}})^{1/3} \sim 1.4\,\mathrm{Mpc}$, a scale above which dark energy dominates. Therefore, from an analytic point of view, the initial parameters of this example sit right at a watershed. In addition, we acknowledge the limitation that the isolated pair idealization is no longer clearly valid after the separation exceeds 1.2\,Mpc, since the surrounding large-scale structure is not accounted for.

In summary, for low and intermediate M31 tangential velocities, Hubble expansion has little effect on the MW-M31 merger. However, for high M31 tangential velocities, Hubble expansion introduces a noticeable delay in the merger time relative to the Newtonian case and can even alter the final fate of the system. Current constraints on the galaxy masses and proper motions are still insufficient to fully determine the exact influence of Hubble expansion on the MW-M31 merger, and future more precise measurements are needed.

The above conclusions can be explained by the competition between the gravitational term $-GM_{\rm total}/r^{2}$, the centrifugal term $l^{2}/r^{3}$ (or equivalently, $V_{\rm tan}^{2}/r$), and the dark energy repulsive term $\frac{1}{3}\Lambda c^2 r$ (or equivalently, $\Omega_\Lambda H_0^{2}r$)~\citep{tanoglidis2015,strigari2025}. The gravitational term is proportional to the mass and inversely proportional to the square of the distance, whereas the dark energy repulsive term is proportional to the distance. Therefore, the effect of expansion is more pronounced for lower mass and higher tangential velocity (i.e., larger post-pericenter apocenter). For example, the post-pericenter apocenter in the upper right panel of Figure~\ref{fig:idealized} is $\sim 300\,\mathrm{kpc}$, for which the ratio of the dark energy repulsive term to the gravitational term is computed to be only $\sim 0.01$, so the effect of the expansion is limited; in the lower right panel, the post-pericenter apocenter (in the Newtonian case) is $\sim 800\,\mathrm{kpc}$, for which the ratio is computed to be $\sim 0.18$, so the expansion significantly affects the outcome.

\subsection{Differences Between Two Expansion Models}
The $w_0w_a$CDM simulations (with high tangential velocities) also exhibit a delay in the merger time relative to the Newtonian ones, but the delay is less pronounced than in the $\Lambda$CDM scenario. In particular, for the lower right panel of Figure~\ref{fig:idealized}, the MW and M31 do not separate but instead still merge in the $w_0w_a$CDM simulation.

For the adopted parameters ($w_{0}=-0.752$, $w_{a}=-0.86$), the equation of state $w(a)=w_{0}+w_{a}(1-a)$ increases with time and crosses zero at $a\sim1.87$. Beyond this point, the dark energy density, which scales as $a^{-3(1+w(a))}$, decays at least as fast as the matter density, and the dark-energy-driven acceleration therefore shuts off rapidly. As a result, the expansion decelerates at late times, and the $w_0w_a$CDM simulations approach the Newtonian ones, consistent with the results described above.

There is a prominent caveat that the $w_0w_a$CDM model we apply is a phenomenological fit calibrated to observational data at $z<2.5$, and its extrapolation forward by several tens of Gyr has no observational support. The late-time suppression of dark energy, and hence the eventual merger of the MW and M31 in this scenario, is therefore a property of the adopted parametrization.

\section{Supplementary Simulations Including M33 and the LMC}
\label{sec:m33lmc}
\citet{vandermarel2019} and \citet{sawala2025} have shown that M33 and the LMC can significantly affect the MW--M31 orbit. Therefore, we rerun the three simulations shown in the lower left panel of Figure~\ref{fig:idealized} with M33 and the LMC included, in order to test the influence of the satellite galaxies on the results.

Although the masses and proper motions of M33 and the LMC also have some uncertainties, our purpose here is only to illustrate the influence of the satellite galaxies, rather than to provide the full possible range of solutions for the future of the LG system. We therefore adopt only one parameter combination for the two satellite galaxies as an example. The adopted parameters are as follows. M33: $M_{\rm vir}=3\times10^{11}\,M_\odot$ \citep{sawala2025}, RA = $01^{\rm h}33^{\rm m}50.9^{\rm s}$, Dec = $+30^{\rm d}39^{\rm m}37^{\rm s}$ \citep{mcconnachie2012}, $D = 859\,\mathrm{kpc}$ \citep{ou2023}, $V_{\rm LOS} = -179.2\,\mathrm{km\,s^{-1}}$ \citep{mcconnachie2012}, $\mu_{\alpha*}=30.6\,\mu\mathrm{as\,yr^{-1}}$, and $\mu_{\delta}=19.0\,\mu\mathrm{as\,yr^{-1}}$ \citep{brunthaler2005,wu2025}; LMC: $M_{\rm vir}=1.5\times10^{11}\,M_\odot$ \citep{sawala2025}, RA = $05^{\rm h}23^{\rm m}34.5^{\rm s}$, Dec = $-69^{\rm d}45^{\rm m}22^{\rm s}$ \citep{mcconnachie2012}, $D = 50\,\mathrm{kpc}$ \citep{pietrzynski2019}, $V_{\rm LOS} = 262.2\,\mathrm{km\,s^{-1}}$ \citep{mcconnachie2012}, $\mu_{\alpha}=1910.0\,\mu\mathrm{as\,yr^{-1}}$, and $\mu_{\delta}=229.0\,\mu\mathrm{as\,yr^{-1}}$ \citep{kallivayalil2013}. M33 and the LMC are modeled as Hernquist halos matched to the NFW profile with concentration $c=12$.

\begin{figure}
    \centering
    \includegraphics[width=1.0\linewidth]{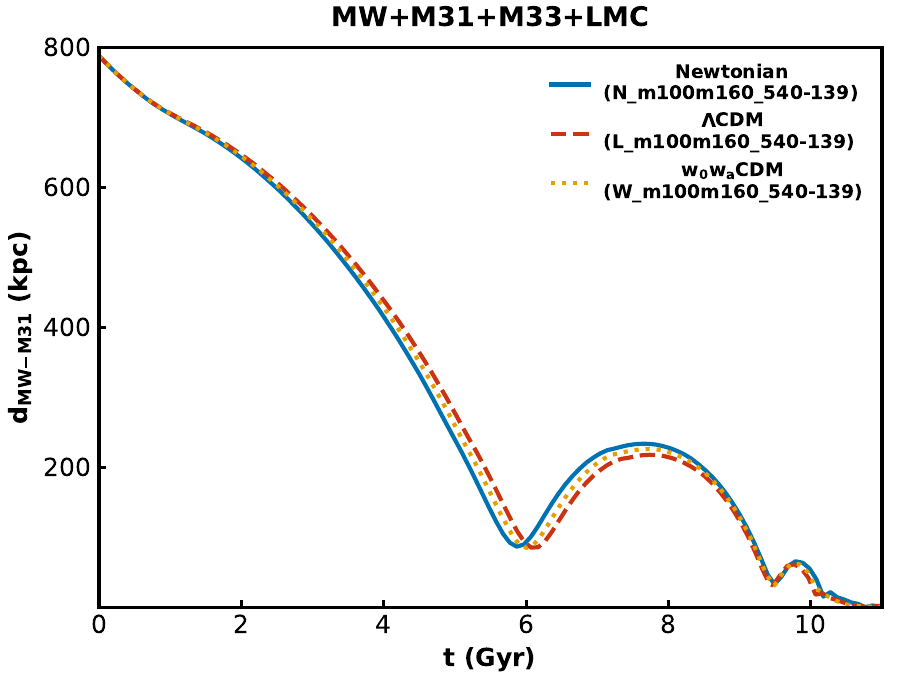}
    \caption{Distance--time curves of the three simulations including M33 and the LMC.}
    \label{fig:idealized_m33}
\end{figure}

The distance--time curves of the three simulations are shown in Figure~\ref{fig:idealized_m33}. Compared with the lower left panel of Figure~\ref{fig:idealized}, the inclusion of the two satellite galaxies delays the time of the first pericenter and reduces the first pericenter distance, consistent with \citet{vandermarel2019}. In addition, as described by \citet{sawala2025}, the presence of M33 and the LMC advances the MW--M31 merger time.

In the simulations including satellite galaxies, the differences in the final merger time among the Newtonian, $\Lambda$CDM, and $w_0w_a$CDM scenarios are reduced. On one hand, this is because the post-pericenter apocenter and the merger time are reduced, which is unfavorable for the expansion to manifest its effect. On the other hand, it is also because the lower pericenter distance makes the MW and M31 suffer stronger energy dissipation by dynamical friction near the first pericenter, and the dissipation is even stronger in the $\Lambda$CDM (or $w_0w_a$CDM) simulations because their pericenter passage occurs later (see point (2) of Section~\ref{subsec:mergertime}). Here, the merger time of the $\Lambda$CDM simulation is indeed slightly earlier than that of the Newtonian one, just like the example in point (2) of Section~\ref{subsec:mergertime} (the upper right panel of Figure~\ref{fig:idealized}).

These three simulations serve only as an example to illustrate the effect of the satellite galaxies; the extent to which the influence of the Hubble expansion on the merger timescale is affected by the satellite galaxies depends on the actual measurements of the galaxy masses and motions.

\section{Conclusion}
\label{sec:conclusion}
We study the differences between galaxy merger simulations with and without Hubble expansion, using the MW--M31 merger as an example.

We first perform controlled idealized simulations containing only the MW and M31, adopting $M_{\rm MW}=8\times10^{11}\,M_{\odot}$ or $1\times10^{12}\,M_{\odot}$, $M_{\rm M31}=1.6\times10^{12}\,M_{\odot}$ or $1\times10^{12}\,M_{\odot}$, proper motion value (along the right ascension direction) of $\mu_{\alpha*}=37.8$, $45.9$, and $54.0\,\mu\mathrm{as\,yr^{-1}}$, and proper motion value (along the declination direction) of $\mu_{\delta}=-27.1$, $-20.5$, and $-13.9\,\mu\mathrm{as\,yr^{-1}}$. All adopted values lie within current observational uncertainties. Each simulation setup is run with \textsc{gadget-4} \citep{springel2021} under Newtonian, $\Lambda$CDM, and $w_0w_a$CDM scenarios, respectively. Then, we perform several simulations including M33 and the LMC to test whether the presence of satellite galaxies can obscure the expansion effect.

Based on our simulation results, we draw the following conclusions:

(1) Hubble expansion causes a delay of the first pericenter passage. For orbits with a large first pericenter distance, it also delays the final merger time; however, for orbits with a small first pericenter distance, it instead advances the final merger time, because of the more prolonged dynamical friction near the first pericenter.

(2) For simulations containing only the MW and M31, the effect of Hubble expansion is negligible at low M31 tangential velocities, especially in cases where the two galaxies merge within a Hubble time. However, at the high tangential velocity ($\gtrsim 66.5\,\mathrm{km\,s^{-1}}$), the $\Lambda$CDM runs introduce a noticeable delay in the merger time compared to the Newtonian ones; in the low galaxy mass cases, the expansion even prevents the two galaxies from merging. But it should be noted that in the simulations where Hubble expansion has a significant effect, the isolated-pair idealization and the assumed expansion history both break down.

(3) The dynamical dark energy model ($w_0w_a$CDM), which is favored by recent DESI DR2 results, also leads to a delay in the merger time relative to the Newtonian case, but the delay is less pronounced than in the $\Lambda$CDM model.

(4) When M33 and the LMC are included, the differences in the MW--M31 merger time between the Newtonian simulations and the $\Lambda$CDM (or $w_0w_a$CDM) simulations are reduced, owing to the decreases in the first pericenter distance and the merger time.

In summary, Hubble expansion can play a noticeable role for galaxy mergers with high tangential velocities (and especially low masses), but current constraints on galaxy masses and proper motions are insufficient to determine its importance in the MW--M31 merger. More precise measurements in the future will help establish whether the expansion effect is negligible or not in this problem.

%% Please use the acknowledgment and contribution environments. This will 
%% be anonomyized when the "anonymous" style option is used. 
\begin{acknowledgments}
We thank Volker Springel for his help regarding the implementation of Hubble expansion in \textsc{gadget-4}, and we also thank Adam Riess and Xuandong Jia for their help in our understanding of the concept of Hubble expansion. We thank Till Sawala for helpful discussions. We thank the anonymous referee for their helpful comments and suggestions.  
This work is supported by the National Natural Science Foundation of China (grant 12225302) and the Fundamental Research Funds for the Central Universities (grant KG202502).
Tianchi Zhang acknowledges support from the National Natural Science Foundation of China (Grants No. 12403008), the Young Data Scientist Program of the China National Astronomical Data Center (Grants No. NADC2024YDS-03), and the science platform of China National Astronomical Data Center.
Shihong Liao acknowledges support by the NSFC grant (Nos. 12588202 and 12473015).
\end{acknowledgments}

\software{GADGET-4 \citep{springel2021}}


\bibliography{hubble}{}
\bibliographystyle{aasjournalv7}

\appendix

\setcounter{figure}{0}
\def\thefigure{A\arabic{figure}}

\section{Robustness tests of the long integrations}
\label{sec:appendix}
Among our 108 MW--M31 merger simulations, the longest one runs for $\sim 60$ Gyr. To demonstrate the robustness of the long integrations, we perform the following tests.

\begin{figure}[h]
    \centering
    \includegraphics[width=0.9\linewidth]{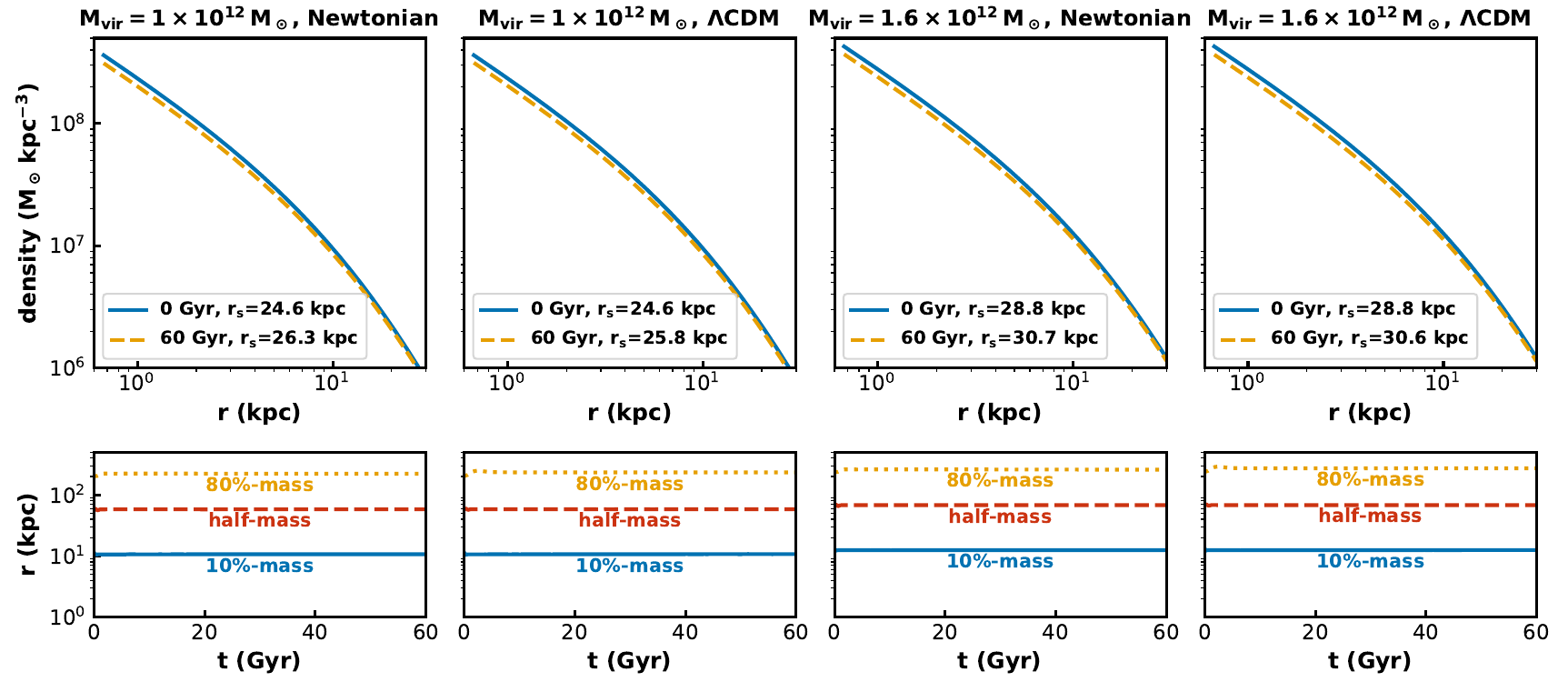}
    \caption{Isolated evolution and stability test of the two fiducial galaxy models in the Newtonian and $\Lambda$CDM cases. Upper panels: dark matter distribution of galaxies at 0 and 60\,Gyr, and the legends give the fitting results of the scale length for the NFW model. Lower panels: time evolution of the Lagrangian radii (10\%-mass, half-mass, and 80\%-mass radii) of galaxies.}
    \label{figs:isolation}
\end{figure}

(1) We evolve the fiducial MW and M31 models in isolation for $60\,\mathrm{Gyr}$ in both the Newtonian and the $\Lambda$CDM setups. Figure~\ref{figs:isolation} shows that the density profiles and the Lagrangian radii are stable at the adopted resolution and softening.

\begin{figure}
    \centering
    \includegraphics[width=0.5\linewidth]{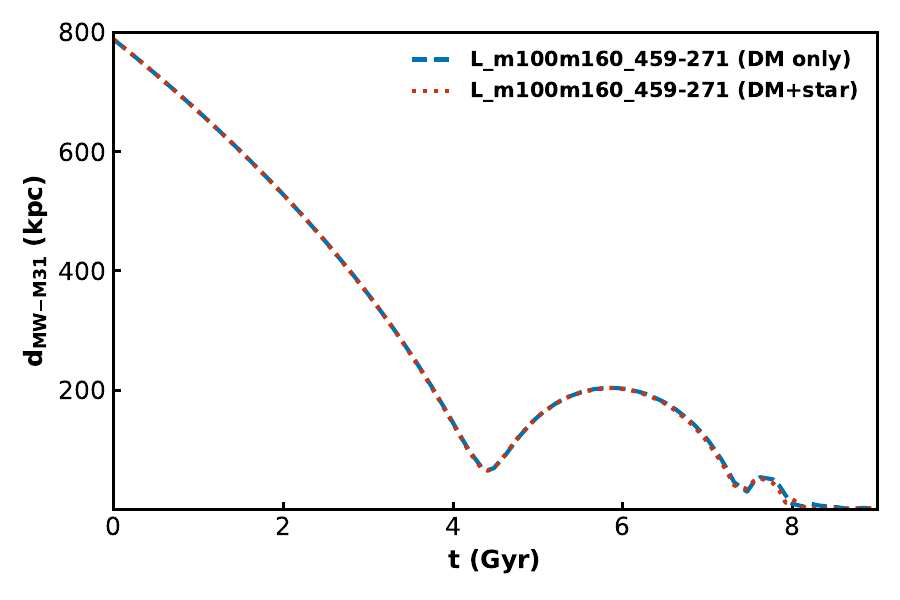}
    \caption{A comparison between a dark-matter-only simulation and its corresponding simulation including the stellar disk and bulge components.}
    \label{figs:withstar}
\end{figure}

(2) Our simulations are dark matter only, so here we instead model the MW and M31 with exponential disks and Hernquist bulges \citep{klypin2002} and rerun one simulation. The MW has a virial mass $M_{\rm vir}=1\times10^{12}\,M_\odot$, halo concentration $c=12$, disk mass $M_{\rm d}=4.1\times10^{10}\,M_\odot$, bulge mass $M_{\rm b}=8\times10^{9}\,M_\odot$, disk scale length $r_{\rm s}=2.1\,\mathrm{kpc}$, disk scale height $h=0.3\,\mathrm{kpc}$, and bulge scale length $a=0.4\,\mathrm{kpc}$. M31 has $M_{\rm vir}=1.6\times10^{12}\,M_\odot$, $c=12$, $M_{\rm d}=7.04\times10^{10}\,M_\odot$, $M_{\rm b}=1.92\times10^{10}\,M_\odot$, disk scale length $r_{\rm s}=2.8\,\mathrm{kpc}$, $h=0.5\,\mathrm{kpc}$, and $a=0.5\,\mathrm{kpc}$. The stellar particle mass and softening length are $5\times10^{5}\,M_\odot$ and $0.5\,\mathrm{kpc}$, respectively. Although the stellar components can affect the central density profile and therefore the efficiency of orbital decay, Figure~\ref{figs:withstar} shows that the two simulations differ only at the final stages, and the overall outcomes are not affected.

\begin{figure}
    \centering
    \includegraphics[width=0.5\linewidth]{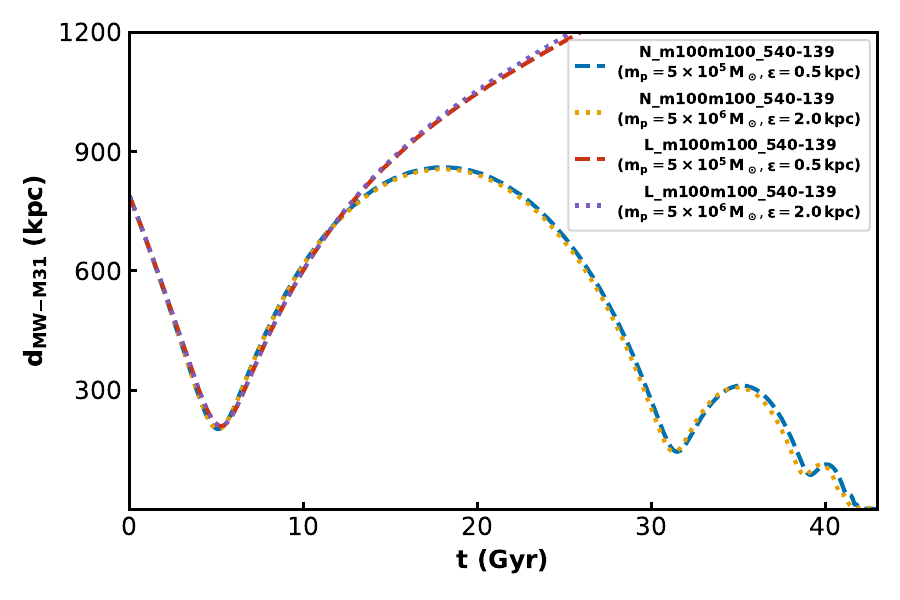}
    \caption{A comparison between two original simulations and their corresponding runs with lower mass resolution and larger softening length.}
    \label{figs:lowresolution}
\end{figure}

(3) Our fiducial particle mass and softening length are $5\times10^{5}\,M_\odot$ and $0.5\,\mathrm{kpc}$; here we rerun two simulations with a particle mass and a softening length of $5\times10^{6}\,M_\odot$ and $2.0\,\mathrm{kpc}$. Figure~\ref{figs:lowresolution} shows that the simulation results are converged.

\begin{figure}
    \centering
    \includegraphics[width=0.5\linewidth]{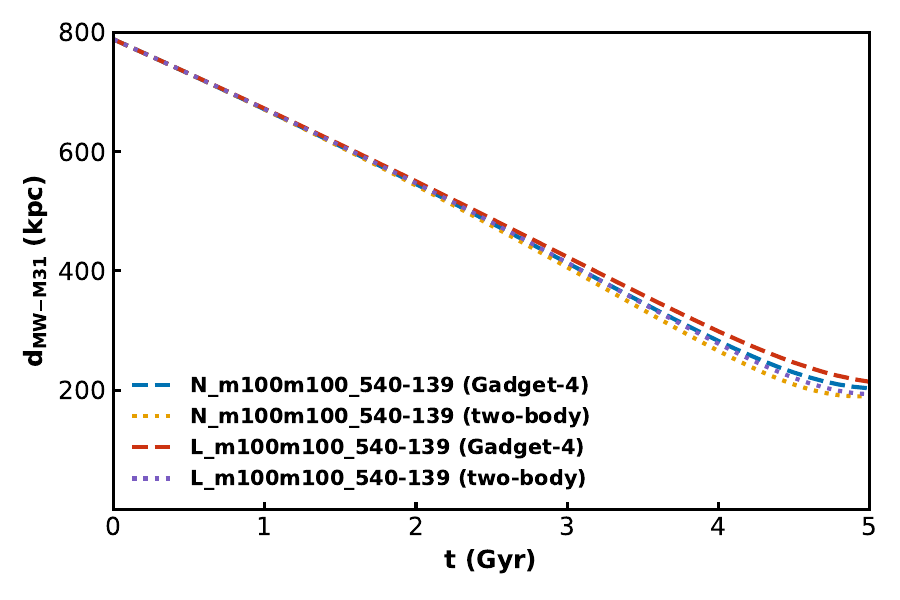}
    \caption{A comparison between two original \textsc{gadget-4} simulations and their corresponding direct two-body integrations.}
    \label{figs:twobody}
\end{figure}

(4) For one set of initial orbital parameters, we treat the MW and M31 as point masses and perform direct two-body integrations using an eighth-order Runge--Kutta method, for the pure Newtonian case and for the case including the dark energy repulsive term, respectively. Figure~\ref{figs:twobody} shows that the \textsc{gadget-4} results agree with the two-body results before the separation drops to $\sim 600\,\mathrm{kpc}$, where the dark matter halos start to overlap and the two-body assumption is no longer valid.

\begin{figure}
    \centering
    \includegraphics[width=0.9\linewidth]{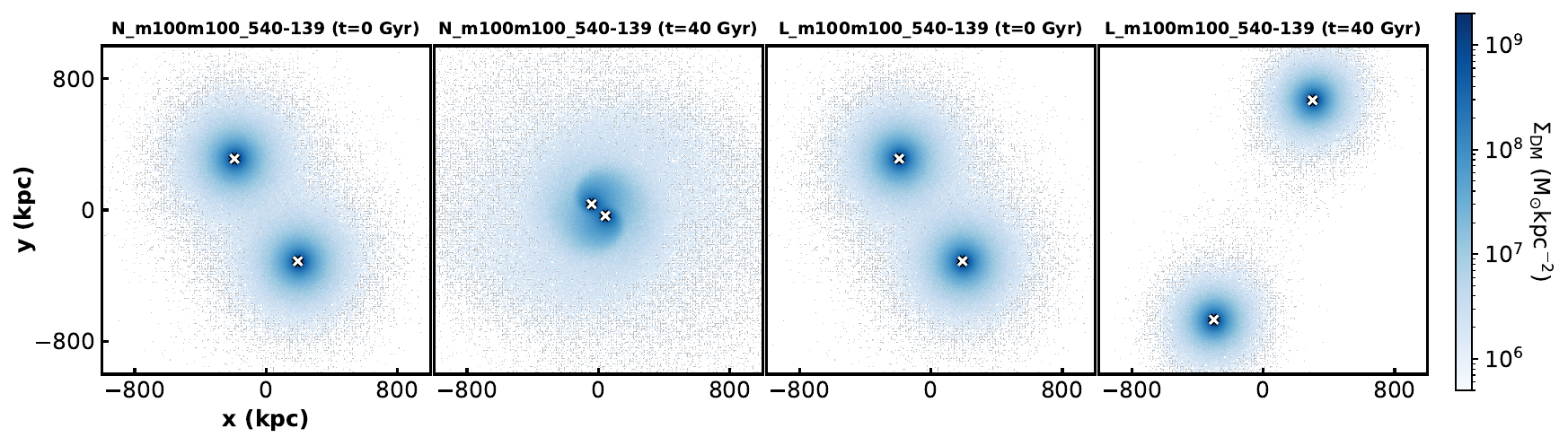}
    \caption{Visualization of the dark matter (blue) and the black hole particle (white cross) in the snapshots at 0 and 40\,Gyr of the Newtonian and $\Lambda$CDM simulations in the lower right panel of Figure~\ref{fig:idealized}.}
    \label{figs:bh}
\end{figure}

(5) We use a black hole particle, whose mass is 100 times the dark matter particle, to mark the galactic center, but it may random walk relative to the true density peak. So we also try to use the center of mass of all dark matter particles of each galaxy to mark the center, and the resulting separations agree with those based on the black hole before the galaxies merge. In addition, we have checked that the black hole particles indeed track the density peak, as shown in Figure~\ref{figs:bh}.

%% For this sample we use BibTeX plus aasjournalv7.bst to generate the



\end{document}